# Three-party secure semiquantum summation without entanglement among quantum user and classical users

Jia-Li, Hu, Tian-Yu Ye*

College of Information & Electronic Engineering, Zhejiang Gongshang University, Hangzhou 310018, P.R.China

E-mail：yetianyu@mail.zjgsu.edu.cn

**Abstract:** In this paper, a three-party secure semiquantum summation protocol, which can calculate the modulo 2 addition of the private bits from one quantum participant and two classical participants, is constructed by only using single qubits as the initial quantum resource. This protocol needs none of quantum entanglement swapping, the unitary operation or a pre-shared private key. This protocol only requires the quantum participant to perform the $\{|0\rangle,|1\rangle\}$ basis measurements, the $\{|+\rangle,|-\rangle\}$ basis measurements and the Bell basis measurements. Compared with the existing only semiquantum summation protocol (Int J Theor Phys, 60 (2021) 3478), this protocol has better performance in quantum measurements for quantum participant; moreover, it may also have higher qubit efficiency.

## 1  Introduction

Secure quantum summation, which is one of the basic problems of quantum secure computation, has been a hot topic of study in recent years. The goal of secure quantum summation is to output the correct summation result of the private inputs from different parties while keeping them secret. A lot good achievements on secure quantum summation [1-15] have been derived from different quantum technologies during the last two decades. Unfortunately, each of the quantum summation protocols in Refs.[3-15] requires all parties to be equipped with full quantum capabilities, which may be unpractical in some circumstances.

Recently, the brand-new concept of semiquantum cryptography was proposed for the first time by Boyer *et al.* [16-17] in their semiquantum key distribution (SQKD) protocols, which permits the classical parties not to have full quantum capabilities. In the years of 2020 and 2022, Ye *et al.* [18-19] employed single photons in both polarization and spatial-mode degrees of freedom to design two novel SQKD protocols. Referring to Refs.[16-17], the classical parties are always restricted to implement the four operations: ①sending the qubits without disturbance; ②measuring the qubits with the $\{|0\rangle,|1\rangle\}$ basis; ③generating the qubits in the $\{|0\rangle,|1\rangle\}$ basis; ④scrambling the qubits. Compared with the conventional quantum cryptography, semiquantum cryptography releases a part of parties from the preparation and measurement of quantum superposition states and quantum entangled states. The counterpart of quantum summation in the realm of semiquantum cryptography is semiquantum summation. In the year of 2021, Zhang *et al.* [20] constructed the first semiquantum summation protocol by using single qubits, which can compute the modulo 2 addition of the private inputs from three classical parties under the help of a quantum semi-honest third party (TP).

Based on the above analysis, in this paper, different from the only existing semiquantum summation protocol of Ref.[20], in order to accomplish the goal of calculating the modulo 2 addition of the private bits from one quantum participant and two classical participants, a three-party secure semiquantum summation protocol is put forward by only using single qubits as the initial quantum resource. Compared with the protocol of Ref.[20], the proposed protocol has better performance in quantum measurements for quantum participant; moreover, it may also have higher qubit efficiency.

## 2  The proposed three-party secure semiquantum summation protocol

The four Bell states can be represented as

$$|\phi^+\rangle = \frac{1}{\sqrt{2}}(|00\rangle + |11\rangle) = \frac{1}{\sqrt{2}}(|+\rangle|+\rangle + |-\rangle|-\rangle), \tag{1}$$

$$|\phi^-\rangle = \frac{1}{\sqrt{2}}(|00\rangle - |11\rangle) = \frac{1}{\sqrt{2}}(|+\rangle|-\rangle + |-\rangle|+\rangle), \tag{2}$$

$$|\psi^+\rangle = \frac{1}{\sqrt{2}}(|01\rangle + |10\rangle) = \frac{1}{\sqrt{2}}(|+\rangle|+\rangle - |-\rangle|-\rangle), \tag{3}$$

$$|\psi^-\rangle = \frac{1}{\sqrt{2}}(|01\rangle - |10\rangle) = \frac{1}{\sqrt{2}}(|+\rangle|-\rangle - |-\rangle|+\rangle), \tag{4}$$

where $|\pm\rangle = (|0\rangle \pm |1\rangle)/\sqrt{2}$. According to Eqs.(1-4), it can be easily obtained that [19]

$$|0\rangle|0\rangle = \frac{1}{\sqrt{2}}(|\phi^+\rangle + |\phi^-\rangle), \tag{5}$$

$$|0\rangle|1\rangle = \frac{1}{\sqrt{2}}(|\psi^+\rangle + |\psi^-\rangle), \tag{6}$$

$$|1\rangle|0\rangle = \frac{1}{\sqrt{2}}(|\psi^+\rangle - |\psi^-\rangle), \tag{7}$$

$$|1\rangle|1\rangle = \frac{1}{\sqrt{2}}(|\phi^+\rangle - |\phi^-\rangle), \tag{8}$$

$$|+\rangle|+\rangle = \frac{1}{\sqrt{2}}(|\phi^+\rangle + |\psi^+\rangle). \tag{9}$$

Alice, Bob and Charlie's private bit strings are denoted as $X = (x_1, x_2, \cdots, x_n)$, $Y = (y_1, y_2, \cdots, y_n)$ and $Z = (z_1, z_2, \cdots, z_n)$, respectively, where $x_i, y_i, z_i \in \{0,1\}$, $i = 1, 2, \ldots, n$. Suppose that Alice is equipped with full quantum capabilities, but both Bob and Charlie only possess limited quantum capabilities. They aim to obtain the modulo 2 addition of their private bit strings. Based on the semiquantum private comparison (SQPC) protocol of Ref.[21], the following protocol is proposed to achieve this goal, where the quantum channels are supposed to be ideal.

**Step 1:** Alice produces a particle sequence of length $2n(8+\lambda+\varepsilon+\gamma)$ whose particles are all in the state of $|+\rangle$. Afterward, she makes these particles form two subsequences of the same length, $P_b = \{p_b^1, p_b^2, \ldots, p_b^{n(8+\lambda+\varepsilon+\gamma)}\}$ and $P_c = \{p_c^1, p_c^2, \ldots, p_c^{n(8+\lambda+\varepsilon+\gamma)}\}$, and then transmits the particles of $P_b$ to Bob and the particles of $P_c$ to Charlie both in the way of one by one. Note that except the first particle, the next one is sent out by Alice only after she receives the previous one. Here, $\lambda, \varepsilon$ are positive integers; $\gamma$ is some parameter bigger than 0; $p_j^t$ is the $t^{th}$ particle of $P_j$, where $j = b, c$ and

$t = 1, 2, \ldots, n(8 + \lambda + \varepsilon + \gamma)$.

**Step 2:** Bob imposes randomly one of the following two actions on each coming particle in $P_b$: returning it to Alice directly (called as the CTRL action) or transmitting the same state as found to Alice after measuring it in the $\{|0\rangle, |1\rangle\}$ basis (called as the SIFT action). After Bob's actions, $P_b$ is changed into $P_b^{'}$.

Charlie also imposes randomly the CTRL action or the SIFT action on each coming particle in $P_c$. After Charlie's actions, $P_c$ is changed into $P_c^{'}$.

**Step 3:** Alice and Bob cooperate together to check whether there is an Eve attacking the particles transmitted between them. Alice randomly selects $n\lambda$ particles from $P_b^{'}$ and publishes Bob the positions of these chosen ones. As for these chosen particles, Bob tells Alice his actions on them and his measurement results on the SIFT particles; then, Alice uses the $\{|+\rangle, |-\rangle\}$ basis to measure the CTRL particles and the $\{|0\rangle, |1\rangle\}$ basis to measure the SIFT particles; afterward, Alice calculates the error rate of the CTRL particles and the error rate of the SIFT particles. Obviously, when there is no Eve, as for the CTRL particles, Alice's measurement results on them should be same as their prepared states; as for the SIFT particles, Alice's measurement results on them should be identical to Bob's measurement results on them. If either of these two kinds of error rates is abnormally high, the protocol will be stopped immediately.

In the meanwhile, Alice randomly selects $n\lambda$ particles from $P_c^{'}$, publishes Charlie the positions of these chosen ones and works together with Charlie to check whether there is an Eve attacking the particles transmitted between them in the way same as above.

**Step 4:** Alice picks out the $l^{\text{th}}$ particle of the remaining $n(8 + \varepsilon + \gamma)$ particles in $P_b^{'}$ and the $l^{\text{th}}$ particle of the remaining $n(8 + \varepsilon + \gamma)$ particles in $P_c^{'}$ to form the $l^{\text{th}}$ particle group, where $l = 1, 2, \ldots, n(8 + \varepsilon + \gamma)$. Then, Alice measures these $n(8 + \varepsilon + \gamma)$ particle groups with the Bell basis. When Alice's measurement result is $|\phi^+\rangle$ or $|\psi^+\rangle$, she announces it to Bob and Charlie directly; and when her measurement result is $|\phi^-\rangle$ or $|\psi^-\rangle$, she announces 'summation' to Bob and Charlie. If the times that Alice announces 'summation' is abnormally low, the protocol will be terminated immediately by Bob and Charlie; otherwise, it will be carried on.

In order to check whether Alice is honest or not, Bob and Charlie randomly pick out $n\varepsilon$ groups from these $n(8 + \varepsilon + \gamma)$ particle groups. For the particle group where both Bob and Charlie took the CTRL actions, according to Eq.(9), no error is detected when Alice's measurement result is $|\phi^+\rangle$ or $|\psi^+\rangle$; and an error is detected when Alice announces 'summation'. For the particle group where both Bob and Charlie took the SIFT actions, when Alice's measurement result is $|\phi^+\rangle$ or $|\psi^+\rangle$, they check whether Alice's measurement result complies with Eqs.(5-8) or not. If Alice is found to be dishonest, the protocol will be halted immediately; otherwise, it will be carried on.

**Step 5:** Among the remaining $n(8 + \gamma)$ particle groups, the number of particle groups where both Bob and Charlie took the SIFT actions and Alice announced 'summation' is $n + \frac{n\gamma}{8}$. Bob and Charlie select the first $n$ particle groups from these $n+\frac{n\gamma}{8}$ ones to compute the summation and announce the positions of these $n$ chosen particle groups to Alice. Then, Alice, Bob and Charlie produce the corresponding one-time pad key bits from their respective measurement results for these $n$ particle groups. When Alice's measurement result on the corresponding particles from the $i^{th}$ particle group is $|\phi^-\rangle$, then $k_i^a=0$ is derived; otherwise, $k_i^a=1$ is obtained. When Bob's measurement result on the corresponding particle from the $i^{th}$ particle group is $|0\rangle$, then $k_i^b=0$ is derived; otherwise, $k_i^b=1$ is obtained. Likewise, when Charlie's measurement result on the corresponding particle from the $i^{th}$ particle group is $|0\rangle$, then $k_i^c=0$ is derived; otherwise, $k_i^c=1$ is obtained. Here, $i=1,2,\ldots,n$. Afterward, Alice, Bob and Charlie compute $r_i^a=k_i^a\oplus x_i$, $r_i^b=k_i^b\oplus y_i$ and $r_i^c=k_i^c\oplus z_i$, respectively, where $\oplus$ is the modulo 2 addition. Here, let $K_A=(k_1^a,k_2^a,\ldots,k_n^a)$, $K_B=(k_1^b,k_2^b,\ldots,k_n^b)$ and $K_C=(k_1^c,k_2^c,\ldots,k_n^c)$. Then, Alice, Bob and Charlie publicly announce $R_A$, $R_B$ and $R_C$, respectively, where $R_A=(r_1^a,r_2^a,\ldots,r_n^a)$, $R_B=(r_1^b,r_2^b,\ldots,r_n^b)$ and $R_C=(r_1^c,r_2^c,\ldots,r_n^c)$. Finally, each of Alice, Bob and Charlie computes $s_i=r_i^a\oplus r_i^b\oplus r_i^c$ and obtains the summation result $S$, where $S=(s_1,s_2,\ldots,s_n)$.

Until now, the description of the proposed three-party semiquantum summation protocol has been finished. It needs to be pointed out that the security check method towards Eve in the proposed protocol is same as that towards Eve in the SQPC protocol of Ref.[21]; and the difference between the honesty check method towards Alice in the proposed protocol and that towards TP in the SQPC protocol of Ref.[21] is that the times that Alice announces 'summation' is checked by Bob and Charlie in the former, but the latter doesn't check it.

## 3  Correctness analysis

In the proposed protocol, Alice, Bob and Charlie want to calculate the modulo 2 addition of $X$, $Y$ and $Z$. Alice, Bob and Charlie utilize the first $n$ particle groups from the $n+\frac{n\gamma}{8}$ ones where both Bob and Charlie took the SIFT actions and Alice announced 'summation' to compute the summation. As a result, it is easily obtained that

$$k_i^a\oplus k_i^b\oplus k_i^c=0, \tag{10}$$

where $i=1,2,\ldots,n$. After Alice, Bob and Charlie publicly announce $R_A$, $R_B$ and $R_C$, respectively,

each of them computes

$$s_i = r_i^a \oplus r_i^b \oplus r_i^c = \left(k_i^a \oplus x_i\right) \oplus \left(k_i^b \oplus y_i\right) \oplus \left(k_i^c \oplus z_i\right)$$

$$= \left(k_i^a \oplus k_i^b \oplus k_i^c\right) \oplus \left(x_i \oplus y_i \oplus z_i\right). \tag{11}$$

After Eq.(10) is inserted into Eq.(11), it has

$$s_i = x_i \oplus y_i \oplus z_i. \tag{12}$$

Consequently, the output of summation is correct.

## 4  Security analysis

### 4.1  Outside attack

The particles in $P_b$ are sent forth and back between Alice and Bob, while the particles in $P_c$ are sent forth and back between Alice and Charlie. Apparently, the transmission mode of the particles in $P_b$ is essentially similar to that of the particles in $P_c$. Without loss of generality, here only consider Eve's attacks on the particles in $P_b$.

(1) Trojan horse attacks

It has been widely accepted that during the round travelling of particles, the particle receiver can adopt a filter to defeat the invisible photon eavesdropping attack and a photon number splitter to prevent the delay-photon Trojan horse attack [22-23].

(2) Double CNOT attack

Eve may try to launch the double CNOT attack [16] as follows: she imposes the CNOT operation on the particle in $P_b$ from Alice and her ancillary particle, where the former and the latter are the control qubit and the target qubit, respectively. After Bob finishes his action, Eve imposes the CNOT operation again on the particle in $P_b'$ from Bob and her ancillary particle, where the former and the latter are also the control qubit and the target qubit, respectively. Ref.[21] has proven in detail that by launching this kind of attack, Eve can only achieve the goal of not being discovered, but she cannot discriminate Bob's action on this particle through her ancillary particle. In other words, she obtains nothing useful about $K_B$ while not being discovered.

(3) Measure-resend attack

Eve intercepts the particles in $P_b$ from Alice to Bob, uses the $\{|0\rangle, |1\rangle\}$ basis to measure them and transmits the fresh ones as found to Bob. With respect to one chosen particle for detecting the existence of Eve, when Bob takes the SIFT action on it, Eve's attack behavior cannot be discovered; and when Bob takes the CTRL action on it, Eve's attack behavior is discovered with the probability of $\frac{1}{2}$; to sum up, Eve's attack behavior is discovered with the probability of $\frac{1}{2} \times \frac{1}{2} = \frac{1}{4}$. For $n\lambda$ chosen particles for detecting the existence of Eve, the probability that Eve's attack behavior is discovered is $1 - \left(\frac{3}{4}\right)^{n\lambda}$. When $nr$ is large enough, this probability is close to 1.

(4) Intercept-resend attack

Eve intercepts the particles in $P_b$ from Alice to Bob and transmits the pre-prepared fake ones in the $\{|0\rangle,|1\rangle\}$ basis to Bob; after Bob's actions, Eve intercepts the ones in $P_b^{'}$ and transmits the ones in $P_b$ to Alice. With respect to one chosen particle for detecting the existence of Eve, when Bob takes the CTRL action on it, Eve's attack behavior cannot be discovered; and when Bob takes the SIFT action on it, Eve's attack behavior is discovered with the probability of $\frac{1}{2}$; to sum up, Eve's attack behavior is discovered with the probability of $\frac{1}{2} \times \frac{1}{2} = \frac{1}{4}$. For $n\lambda$ chosen particles for detecting the existence of Eve, the probability that Eve's attack behavior is discovered is $1-\left(\frac{3}{4}\right)^{n\lambda}$.

## 4.2 Participant attack

We further consider the participant attack, which was firstly put forward by Gao *et al.* [24] in the year of 2007. Apparently, in the proposed protocol, when any two dishonest users collude together, they can deduce out the left one's private bit string from the summation result without any effort. As a result, the proposed protocol cannot resist the collusion attack from any two dishonest users. Consequently, we only need to consider the participant attack from one dishonest user.

(1) Participant attack from Alice

Alice may try her best to extract $Y$ from $R_B$ and $Z$ from $R_C$. In order to achieve these goals, Alice first needs to know $K_B$ and $K_C$, respectively.

**Attack I:** In order to directly get $K_B$ and $K_C$, Alice produces all particles in the $\{|0\rangle,|1\rangle\}$ basis in Step 1 and then sends them to Bob and Charlie, respectively; moreover, Alice always plays no tricks on her announcements to Bob and Charlie in Step 4 with regard to her Bell basis measurement results on the remaining $n(8+\varepsilon+\gamma)$ particle groups. Fortunately, Alice's attack can be detected by the security check from Bob and Charlie in Step 4. For one chosen particle group for checking Alice's honesty, by virtue of Eqs.(5-9), when both Bob and Charlie take the CTRL actions on the corresponding particles, the probability that Alice's attack can be detected is $\frac{1}{2}$; when both Bob and Charlie take the SIFT actions on the corresponding particles, the probability that Alice's attack can be detected is 0; when Bob and Charlie take different actions on the corresponding particles, the probability that Alice's attack can be detected is 0, as Bob and Charlie do not check this situation. As a result, for one chosen particle group for checking Alice's honesty, the probability that Alice's attack can be detected is $\frac{1}{2} \times \frac{1}{2} \times \frac{1}{2} = \frac{1}{8}$. For $n\varepsilon$ chosen particle groups for checking Alice's honesty, Alice's attack can be detected with the probability of $1-\left(\frac{7}{8}\right)^{n\varepsilon}$. When $n\varepsilon$ is large enough, this probability is close to 1.

**Attack II:** Alice uses the $Z$ basis rather than the Bell basis to measure all particles in the remaining $n(8+\varepsilon+\gamma)$ particle groups, and tries to pass the honesty check toward her by announcing the fake Bell basis measurement results to Bob and Charlie in Step 4. For one chosen particle group for checking Alice's honesty, according to Eqs.(5-8), when Alice's measurement result is $|0\rangle|0\rangle$ or $|1\rangle|1\rangle$, she always announces the fake measurement result $|\phi^+\rangle$; and when Alice's measurement result is $|0\rangle|1\rangle$ or $|1\rangle|0\rangle$, she always announces the fake measurement result $|\psi^+\rangle$; as a result, no matter what actions Bob and Charlie take on the corresponding particles, the probability that Alice's attack can be detected is always 0. Although Alice can pass the honesty check toward her, her attack makes the protocol automatically terminated by Bob and Charlie, as the times that she announces 'summation' is 0. In order to let the protocol continued, for one chosen particle group for checking Alice's honesty, Alice randomly announces the fake measurement result $|\phi^+\rangle$ or 'summation', when her measurement result is $|0\rangle|0\rangle$ or $|1\rangle|1\rangle$; and she randomly announces the fake measurement result $|\psi^+\rangle$ or 'summation', when her measurement result is $|0\rangle|1\rangle$ or $|1\rangle|0\rangle$; as a result, when both Bob and Charlie take the SIFT actions on the corresponding particles, the probability that Alice's attack can be detected is 0; and when both Bob and Charlie take the CTRL actions on the corresponding particles, the probability that Alice's attack can be detected is $\frac{1}{2}$. Hence, for one chosen particle group for checking Alice's honesty, Alice's attack can be detected with the probability of $\frac{1}{2} \times \frac{1}{2} \times \frac{1}{2} = \frac{1}{8}$. For $n\varepsilon$ chosen particle groups for checking Alice's honesty, Alice's attack can be detected with the probability of $1-\left(\frac{7}{8}\right)^{n\varepsilon}$.

(2) Participant attack from Bob or Charlie

In the proposed protocol, Bob plays the same role as Charlie. Without loss of generality, here assume that Bob is dishonest. Bob may try his best to extract $X$ from $R_A$ and $Z$ from $R_C$. In order to achieve these goals, Bob first needs to know $K_A$ and $K_C$, respectively. $K_A$ and $K_C$ are derived from the particle groups where both Bob and Charlie took the SIFT actions and Alice announced 'summation'. Bob may try to implement the following attack to get $K_C$: he intercepts and uses the $\{|0\rangle,|1\rangle\}$ basis to measure the corresponding particles in $P_c$ sent out from Alice whose counterparts in $P_b$ were performed with the SIFT operations by him, and transmits the resulted particles to Charlie. However, Bob has no access to Charlie's actions on them, which results in his cheating behavior being discovered. With respect to one chosen position for detection in Step 3, the probability that Bob took the SIFT action on the corresponding particle in $P_b$ is $\frac{1}{2}$; the probability that Charlie took the CTRL action on the corresponding particle in $P_c$ is also $\frac{1}{2}$; as a result, Alice obtains the wrong measurement result $|-\rangle$ with the probability of $\frac{1}{2}$; hence, the probability that Bob's cheating behavior can be discovered in Step 3 is $\frac{1}{2} \times \frac{1}{2} \times \frac{1}{2} = \frac{1}{8}$. For $n\lambda$ chosen positions for detection in Step 3, the probability that Bob's cheating behavior can be detected is $1 - \left(\frac{7}{8}\right)^{nr}$. In addition, because Bob cannot obtain $K_C$ without being discovered, he has no chance to get $K_A$ via $K_B$ and $K_C$. Moreover, Bob cannot obtain $K_A$ through Alice's announcement of 'summation' either.

It can be concluded that Bob has no chance to get $X$ and $Z$ without being discovered.

## 5  Discussions

The qubit efficiency, defined as [25]

$$\eta = \frac{v}{q+f}, \qquad (13)$$

can be employed to evaluate the communication efficiency of a quantum cryptography protocol, where $v$, $q$ and $f$ are the number of private bits for summation, the number of consumed qubits and the number of classical bits consumed for the classical communication, respectively. Here, the classical bits consumed for security check towards Eve and honesty test towards Alice are disregarded.

In this protocol, the private bit string from each of Alice, Bob and Charlie has $n$ bits, so $v = n$; Alice needs to prepare $2n(8+\lambda+\varepsilon+\gamma)$ $|+\rangle$, while each of Bob and Charlie needs to produce $\frac{n(8+\lambda+\varepsilon+\gamma)}{2}$ qubits in the $\{|0\rangle, |1\rangle\}$ basis for the SIFT actions, so $q = 2n(8+\lambda+\varepsilon+\gamma) + \frac{n(8+\lambda+\varepsilon+\gamma)}{2} \times 2 = 3n(8+\lambda+\varepsilon+\gamma)$; and Alice, Bob and Charlie publicly announce $R_A$, $R_B$ and $R_C$ via the classical channel, respectively, so $f = 3n$. Hence, the qubit efficiency of this protocol is $\eta = \frac{n}{3n(8+\lambda+\varepsilon+\gamma)+3n} = \frac{1}{3(8+\lambda+\varepsilon+\gamma)+3}$.

In the protocol of Ref.[20], the private bit string from $P_j$ has $n$ bits, where $j = 1,2,3$, so $v = n$; TP needs to prepare $3n(32+r+d+\delta)$ $|+\rangle$, while $P_j$ needs to produce $\frac{n(32+r+d+\delta)}{2}$ qubits in the $\{|0\rangle, |1\rangle\}$ basis for the SIFT actions, so $q = 3n(32+r+d+\delta) + \frac{n(32+r+d+\delta)}{2} \times 3 = \frac{9n(32+r+d+\delta)}{2}$; and $P_j$ needs to send $C_j$ to TP via the classical channel, so $f = 3n$. Hence, the qubit efficiency of the protocol of Ref.[18] is

$\eta = \frac{n}{\frac{9n(32+r+d+\delta)}{2}+3n} = \frac{2}{9(32+r+d+\delta)+6}$.

According to Table 1, the differences between this protocol and the one of Ref.[20] can be summarized as follows: (1) this protocol aims to calculate the summation of the private bits from one quantum participant and two classical participants, while the one of Ref.[20] is to compute the summation of the private bits from three classical participants; (2) there are three participants involved totally in this protocol, while there are four participants involved totally in the one of Ref.[20]; (3) this protocol needs the $\{|0\rangle,|1\rangle\}$ basis measurements, the $\{|+\rangle,|-\rangle\}$ basis measurements and the Bell basis measurements for quantum participant, while the one of Ref.[20] needs the $\{|0\rangle,|1\rangle\}$ basis measurements, the $\{|+\rangle,|-\rangle\}$ basis measurements and the GHZ-type basis measurements for quantum participant; as a result, this protocol exceeds the one of Ref.[20] in quantum measurements for quantum participant, as the Bell basis measurements are much easier to realize in practice than the GHZ-type basis measurements; (4) when the parameters $\lambda$, $\varepsilon$ and $\gamma$ are suitably chosen (as long as $2(8+\lambda+\varepsilon+\gamma)<3(32+r+d+\delta)$), this protocol can easily defeat the one of Ref.[20] in the qubit efficiency.

Table 1 Comparison results of different semiquantum summation protocols

| | The protocol of Ref.[20] | This protocol |
|---|---|---|
| Quantum channel | ideal quantum channel | ideal quantum channel |
| Function | calculating the summation of the private bits from three classical participants | calculating the summation of the private bits from one quantum participant and two classical participants |
| Characteristic | measure-resend | measure-resend |
| Number of involved participants | four | three |
| Initial quantum resource | $|+\rangle$ | $|+\rangle$ |
| Quantum measurement of quantum participant | $\{|0\rangle,|1\rangle\}$ basis measurements, $\{|+\rangle,|-\rangle\}$ basis measurements and GHZ-type basis measurements | $\{|0\rangle,|1\rangle\}$ basis measurements, $\{|+\rangle,|-\rangle\}$ basis measurements and Bell basis measurements |
| Quantum measurement of classical participants | $\{|0\rangle,|1\rangle\}$ basis measurements | $\{|0\rangle,|1\rangle\}$ basis measurements |
| Usage of quantum entanglement swapping | No | No |
| Usage of unitary operations | No | No |
| Usage of pre-shared private key | No | No |
| Summation type | modulo 2 addition | modulo 2 addition |
| Computation way | bit-by-bit | bit-by-bit |
| Qubit efficiency | $\frac{2}{9(32+r+d+\delta)+6}$ | $\frac{1}{3(8+\lambda+\varepsilon+\gamma)+3}$ |

## 6 Conclusions

In summary, this paper proposes a three-party secure semiquantum summation protocol without entanglement which can calculate the modulo 2 addition of the private bits from one quantum participant and two classical participants. The proposed protocol adopts none of quantum entanglement swapping, the unitary operation or a pre-shared private key. Because semiquantum summation is a brand-new branch of semiquantum cryptography, we will put more energies into studying it in the future. For instance, it is worthy of studying how to make a semiquantum summation protocol resistant against different kinds of noise, how to enhance its qubit efficiency, and so on.

## Acknowledgments

Funding by the National Natural Science Foundation of China (Grant No.62071430) and the Fundamental Research Funds for the Provincial Universities of Zhejiang (Grant No.JRK21002) is gratefully acknowledged.